\documentclass[conference]{IEEEtran}
\IEEEoverridecommandlockouts
\usepackage{cite}
\usepackage{amsmath,amssymb,amsfonts}
\usepackage{algorithm}  
\usepackage{algpseudocode}
\usepackage{multirow}

\usepackage{amsfonts} 

\usepackage{hyperref}
\usepackage{subcaption}
\usepackage{array} 
\usepackage{makecell} 
\usepackage[flushleft]{threeparttable}
\usepackage{tabularx} 
\usepackage{graphicx}
\usepackage{epstopdf}

\usepackage{textcomp}
\usepackage{xcolor}
\def\BibTeX{{\rm B\kern-.05em{\sc i\kern-.025em b}\kern-.08em
    T\kern-.1667em\lower.7ex\hbox{E}\kern-.125emX}}
\def\BibTeX{{\rm B\kern-.05em{\sc i\kern-.025em b}\kern-.08em
    T\kern-.1667em\lower.7ex\hbox{E}\kern-.125emX}}
    \title{Adaptive Field Effect Planner for Safe Interactive Autonomous Driving on Curved Roads\\
 \thanks{$^{\dagger}$ Common contribution}
 \thanks{$^{\S}$Corresponding author}%
}

\author{
\IEEEauthorblockN{1\textsuperscript{st} Qinghao Li$^{\dagger}$}
\IEEEauthorblockA{\textit{Department of Computer Science,} \\
\textit{University of Liverpool,}\\
Liverpool L69 3GJ, United Kingdom \\
psqli35@liverpool.ac.uk}
\and
\IEEEauthorblockN{2\textsuperscript{nd} Zhen Tian$^{\dagger}$}
\IEEEauthorblockA{\textit{James Watt School of Engineering,} \\
\textit{University of Glasgow,}\\
Glasgow G12 8QQ, United Kingdom \\
2620920z@student.gla.ac.uk}
\and
\IEEEauthorblockN{3\textsuperscript{th} Xiaodan Wang}
\IEEEauthorblockA{\textit{School of Engineering,} \\
\textit{Cardiff University,}\\
Cardiff CF24 3AA, United Kingdom \\
WangX223@cardiff.ac.uk}
\and
\IEEEauthorblockN{4\textsuperscript{th} Jinming Yang}
\IEEEauthorblockA{\textit{School of Computing Science,} \\
\textit{University of Glasgow,}\\
Glasgow G12 8QQ, United Kingdom \\
j.yang.8@research.gla.ac.uk}
\and
\IEEEauthorblockN{5\textsuperscript{th} Zhihao Lin$^{\S}$}
\IEEEauthorblockA{\textit{James Watt School of Engineering,} \\
\textit{University of Glasgow,}\\
Glasgow G12 8QQ, United Kingdom \\
2800400L@student.gla.ac.uk}
}

\begin{document}
\maketitle
\begin{abstract}
Autonomous driving has garnered significant attention for its potential to improve safety, traffic efficiency, and user convenience. However, the dynamic and complex nature of interactive driving poses significant challenges, including the need to navigate non-linear road geometries, handle dynamic obstacles, and meet stringent safety and comfort requirements. Traditional approaches, such as artificial potential fields (APF), often fall short in addressing these complexities independently, necessitating the development of integrated and adaptive frameworks. This paper presents a novel approach to autonomous vehicle navigation that integrates artificial potential fields, Frenet coordinates, and improved particle swarm optimization (IPSO). A dynamic risk field, adapted from traditional APF, is proposed to ensure interactive safety by quantifying risks and dynamically adjusting lane-changing intentions based on surrounding vehicle behavior. Frenet coordinates are utilized to simplify trajectory planning on non-straight roads, while an enhanced quintic polynomial trajectory generator ensures smooth and comfortable path transitions. Additionally, an IPSO algorithm optimizes trajectory selection in real time, balancing safety and user comfort within a feasible input range. The proposed framework is validated through extensive simulations and real-world scenarios, demonstrating its ability to navigate complex traffic environments, maintain safety margins, and generate smooth, dynamically feasible trajectories.
\end{abstract}


\section{Introduction}

Autonomous driving technology has garnered significant public interest due to its potential to revolutionize transportation~\cite{parekh2022review,9802527,lin2024conflicts}. By promising to enhance road safety, improve traffic efficiency, and provide greater convenience for users, autonomous vehicles (AVs) are poised to address critical challenges in modern mobility systems~\cite{BADUE2021113816,lin2024enhanced,inproceedings}. The societal benefits, ranging from reducing traffic accidents caused by human error to alleviating urban congestion, have made autonomous driving an important technological frontier use to their advantages in perception~\cite{zhu2025fdnet,lin2025slam2,zhu2024podb}, decision-making~\cite{tian2024efficient} and control.

However, the complexity of interactive driving in dynamic traffic environments, such as highway, on-ramping merging, and roundabouts, presents formidable challenges~\cite{10579912,tian2025evaluating}. This complexity is compounded by the need to balance multiple objectives while adhering to real-time decision-making constraints~\cite{10268996}, together with robust, adaptive responses to HDVs~\cite{martinez2017driving}.

The primary challenges in autonomous navigation arise from two critical aspects: dynamic obstacles and curvy road geometries~\cite{aizat2023comprehensive,chen2023emergency}. Dynamic obstacles demand precise risk assessment and safe decision-making. Simultaneously, non-straight road geometries require AVs to generate road-aligned paths to minimize unnecessary deviations.

In addition to these operational challenges, AV users have high expectations for safety and comfort. Safety entails avoiding collisions and maintaining appropriate distances from obstacles~\cite{nahata2021assessing}, while comfort demands smooth trajectories that minimize abrupt accelerations~\cite{de2023standards}. Balancing safety and comfort underpins the design of effective autonomous driving systems.

To address these challenges, this work integrates three complementary approaches: artificial potential fields (APF) for interactive safety, Frenet coordinates for handling non-straight roads, and quintic polynomial curves for ensuring comfort. APF provide a natural framework for capturing the interactive dynamics between the AV and surrounding vehicles, ensuring safe navigation in complex traffic scenarios~\cite{pan2021improved}. However, traditional APF is not capable of addressing interactive driving on its own and is usually used as a risk-quantification module to assist in decision-making~\cite{hang2020integrated}. Therefore, an adaptation of the traditional APF is required to ensure safe interactive driving. Frenet coordinates allow for trajectory generation in road-aligned coordinates, simplifying optimization on curved roads~\cite{wang2023research,yu2023lf}.

Our contributions build upon these foundations by introducing novel adaptations to each component. First, we adapt the APF into a risk field model that not only ensures safety and efficiency by enabling the AV to adjust its lane-changing intentions based on surrounding HDVs. Second, the Frenet-based trajectory generation framework is utilized to handle curvy road geometries effectively. Third, we employ an adapted quintic polynomial trajectory generator to ensure smoothness for comfort. These adaptations enable a robust and integrated solution to the challenges of autonomous navigation.

This work provides the following contributions to the field of autonomous driving:

\begin{enumerate}
    \item 
    We introduce a novel risk field model based on artificial potential fields, enabling the AV to dynamically adjust its lane-changing intentions by considering the behaviors and positions of surrounding vehicles.

    \item 
    By leveraging Frenet coordinates, we effectively generate road-aligned trajectories that accommodate complex road geometries, including lane-changing scenario, and maneuvers scenario with multiple HDVs on curvy road.

    \item 
    An IPSO-enhanced quintic polynomial curve is used to ensure trajectory smoothness, incorporating a cost function that selects the optimal curve within control input limits, balancing comfort and feasibility. Simulation results suggest that convergence speed and computation efficiency of the IPSO-enhanced quintic polynomial curve is higher than other popular benchmark algorithms.
\end{enumerate}

The rest of the paper is organized as follows: Section II introduces the interactive environment and Frenet Formulation. Section III presents risk field formulation. Section IV presents the improved particle swarm optimization. Section V reports the simulation results. Section VI draws the conclusions.

\section{interactive environment and Frenet Formulation}
\subsection{Interactive traffic environment}
The considered problem includes a inner lane and a outer lane. Forexample, the host vehicle (HV) is positioned in the inner lane, while proceeding vehicle (PV), immediate vehicle (IV), and rear vehicle (RV) occupy the front area of inner lane, front area of outer lane, and rear area of outer lane respectively. Among the whole validation, In this paper the involved vehicles are defined as: the SV in red, the PV in blue, the IV in purple, and the RV in green.

\subsection{State Representation}
The complete state of the system is as follows:
\begin{equation}
    \mathbf{\Xi}(t) = \begin{bmatrix}
    \mathbf{X}_\text{ego}(t) & \mathbf{X}_\text{front}(t) & \mathbf{X}_\text{rear}(t) & \mathbf{X}_\text{adjacent}(t)
    \end{bmatrix}
\end{equation}
where $\mathbf{X}_\text{ego}(t)$ is the ego vehicle state vector, $\mathbf{X}_\text{front}(t)$ is the state of the vehicle directly ahead in the same lane, $\mathbf{X}_\text{rear}(t)$ is the state of the following vehicle, $\mathbf{X}_\text{adjacent}(t)$ denotes the states of vehicles in adjacent lanes that might affect lane-change decisions.

\subsection{Individual Vehicle State}
Each vehicle's state vector is defined as:
\begin{equation}
    \mathbf{X}_i(t) = \begin{bmatrix}
    x_i(t) & y_i(t) & \theta_i(t) & v_i(t) & a_i(t) & \kappa_i(t) & \dot{\psi}_i(t)
    \end{bmatrix}^T
\end{equation}
where $(x_i, y_i)$ is the position in global coordinates, $\theta_i$ is the heading angle relative to the global reference frame, $v_i$ is the longitudinal velocity, $a_i$ describes the acceleration, $\kappa_i$ is the path curvature, and $\dot{\psi}_i$ is the yaw rate capturing the rotational motion of the vehicle around its vertical axis.

\subsection{Frenet Coordinate Transformation}

The navigation and trajectory planning system employs Frenet coordinates to simplify path planning along curved roads. The Frenet frame is defined by two components: the arc length along the reference path $s$, and the lateral offset from the path $d$. This relationship can be expressed as:

\begin{equation}
    \begin{bmatrix}
    s \\
    d
    \end{bmatrix} = \mathbf{T}(x, y)
\end{equation}

where $\mathbf{T}$ represents the transformation from global coordinates $(x, y)$ to Frenet coordinates $(s, d)$. The transformation considers the reference path defined by waypoints:

\begin{equation}
    \mathcal{P}_{ref} = \{(x_i, y_i) | i = 1, ..., N\}
\end{equation}

\subsubsection{Global to Frenet Transformation}
For any point $(x, y)$ in global coordinates, the Frenet coordinates are computed through:

\begin{equation}
    \begin{aligned}
    s &= \int_0^{\xi} \sqrt{\left(\frac{dx}{d\tau}\right)^2 + \left(\frac{dy}{d\tau}\right)^2} d\tau \\
    d &= \text{sign}(\dot{x}\ddot{y} - \ddot{x}\dot{y})\sqrt{(x-x_p)^2 + (y-y_p)^2}
    \end{aligned}
\end{equation}

where $(x_p, y_p)$ is the nearest point on the reference path, and $\xi$ is the path parameter at this point. The sign term ensures the lateral offset is positive when the point is to the left of the path and negative when to the right.

\subsubsection{Dynamic State Transformation}
The complete state transformation including velocities and accelerations is:

\begin{equation}
    \begin{bmatrix}
    \dot{s} \\
    \dot{d} \\
    \ddot{s} \\
    \ddot{d}
    \end{bmatrix} = 
    \begin{bmatrix}
    \frac{\dot{x}\cos\psi + \dot{y}\sin\psi}{1-\kappa d} \\
    \dot{y}\cos\psi - \dot{x}\sin\psi \\
    \frac{\ddot{x}\cos\psi + \ddot{y}\sin\psi}{1-\kappa d} + \frac{\kappa\dot{s}^2}{1-\kappa d} \\
    \ddot{y}\cos\psi - \ddot{x}\sin\psi - \kappa\dot{s}^2
    \end{bmatrix}
\end{equation}

where $\psi$ is the reference path heading, and $\kappa$ is the path curvature at point $s$.

\subsubsection{Frenet to Global Transformation}
The inverse transformation from Frenet to global coordinates is computed as:

\begin{equation}
    \begin{bmatrix}
    x \\
    y \\
    \theta
    \end{bmatrix} =
    \begin{bmatrix}
    x_p - d\sin\psi \\
    y_p + d\cos\psi \\
    \psi + \arctan\left(\frac{\dot{d}}{\dot{s}(1-\kappa d)}\right)
    \end{bmatrix}
\end{equation}

where $(x_p, y_p)$ corresponds to the reference path point at arc length $s$, and the velocities are transformed according to:

\begin{equation}
    \begin{bmatrix}
    \dot{x} \\
    \dot{y}
    \end{bmatrix} =
    \begin{bmatrix}
    \cos\psi & -\sin\psi \\
    \sin\psi & \cos\psi
    \end{bmatrix}
    \begin{bmatrix}
    \dot{s}(1-\kappa d) \\
    \dot{d}
    \end{bmatrix}
\end{equation}

\section{Risk Field Formulation}

\subsection{Attraction Risk Field}
The lane-keeping attraction field is defined as:
\begin{equation}
    U_a(\mathbf{X}_{ego}, t) = \alpha \left(\frac{R}{R_1}\right)^2(2\pi-\theta)^2 + \beta\|\mathbf{X}_\text{ego} - \mathbf{X}_\text{ref}\|_{\mathbf{W}}^2
\end{equation}
where $\alpha$ represents the attraction field strength coefficient, $R$ denotes the current radius of curvature of the vehicle's path, $R_1$ is the reference radius, $\mathbf{X}_\text{ref}$ specifies the reference trajectory, and $\mathbf{W}$ is a positive definite weighting matrix that balances the importance of different state variables.

\subsection{Front Vehicle Repulsive Field Formulation}

The repulsive field generated by the front vehicle serves as a crucial safety mechanism in autonomous driving, created through a sophisticated combination of angular and spatial relationships between vehicles. This field is mathematically expressed through a comprehensive potential function:
\begin{equation}
\begin{aligned}
    U_b(\mathbf{X}_{ego}, \mathbf{X}_{front}, t) = 
    & \, \gamma \left(\frac{\Delta\theta}{\theta_{front} - \theta_{ego}}\right) \cdot \\
    & \, \exp\left(-\frac{\|\mathbf{X}_{ego} - \mathbf{X}_{front}\|^2}{2\sigma^2}\right)
\end{aligned}
\end{equation}
where $\gamma$ is the fundamental repulsion strength coefficient that scales the overall magnitude of the repulsive force, crucially determining how strongly vehicles repel each other when their paths converge. The angular component $\frac{\Delta\theta}{(\theta_\text{front}-\theta_{ego})}$ captures the critical angular relationship between vehicles, with $\Delta\theta$ establishing a safety threshold in the angular domain and $(\theta_\text{front}-\theta_\text{ego})$ measuring the actual angular separation between vehicles. This ratio intensifies the repulsive effect as the angular separation approaches unsafe levels. The exponential term $\exp\left(-\frac{\|\mathbf{X}_\text{ego} - \mathbf{X}_\text{front}\|^2}{2\sigma^2}\right)$ modulates the field strength based on the Euclidean distance between vehicles, where $\sigma$ acts as a spatial decay parameter controlling the rate at which repulsion diminishes with distance. The distance $\|\mathbf{X}_\text{ego} - \mathbf{X}_\text{front}\|$ represents the instantaneous separation between vehicle positions, computed as $\sqrt{(x_\text{ego}-x_\text{front})^2 + (y_\text{ego}-y_\text{front})^2}$. The repulsive field naturally induces trajectory modifications through its gradient:
\begin{equation}
\begin{aligned}
    \mathbf{F}_{repulsive} = -\nabla U_B = & -\gamma \nabla \Bigg[ 
    \left(\frac{\Delta\theta}{\theta_\text{front} - \theta_\text{ego}}\right) \cdot \\
    & \exp\left(-\frac{\|\mathbf{X}_\text{ego} - \mathbf{X}_\text{front}\|^2}{2\sigma^2}\right) 
    \Bigg]
\end{aligned}
\end{equation}

The resulting force $\mathbf{F}_\text{repulsive}$ guides the ego vehicle away from potentially dangerous configurations, with its magnitude adjusting based on the immediacy of the threat. The parameter $\gamma$ can be dynamically adjusted based on vehicle speeds and road conditions:
\begin{equation}
    \gamma = \gamma_0\left(1 + \alpha\frac{v_\text{ego}}{v_\text{safe}} + \beta\frac{v_\text{front}}{v_\text{safe}}\right)
\end{equation}
where $\gamma_0$ is the baseline repulsion strength, $v_\text{safe}$ is a reference safe velocity, and $\alpha, \beta$ are weighting coefficients that modify the repulsion based on the velocities of both vehicles. Similarly, the spatial decay parameter $\sigma$ can be velocity-dependent:

\begin{equation}
    \sigma = \sigma_0\left(1 + \lambda\frac{\max(v_\text{ego}, v_\text{front})}{v_\text{safe}}\right)
\end{equation}

This comprehensive repulsive field formulation integrates seamlessly with other potential fields in the system, such as the lane-keeping attraction field $U_A$ and the lane-change potential field $U_C$, to create a complete navigation strategy that balances safety, efficiency, and comfort. The resulting total potential field $U_\text{total} = U_A + U_B + U_C$ guides the vehicle through complex traffic scenarios while maintaining appropriate safety margins and enabling smooth transitions between different driving behaviors.

\subsection{Lane Change Risk Field}
The lane change potential field is defined as:
\begin{equation}
    U_c(\mathbf{\Xi}, t) = \lambda\left|\frac{v_\text{ego}-\max(v_\text{rear},v_\text{adj})+\xi\Delta\theta}{(v_\text{adj}-v_\text{rear})+\xi(\theta_\text{adj}-\theta_\text{rear})}\right| \cdot \Phi(\mathbf{\Xi})
\end{equation}

This complex field orchestrates safe lane changes through multiple interacting parameters: $\lambda$ serves as the lane change field strength coefficient, $\xi$ acts as the angular velocity scaling factor that weights the importance of angular differences between vehicles, $v_\text{ego}$, $v_\text{rear}$, and $v_\text{adj}$ represent the velocities of the ego vehicle, rear vehicle, and adjacent lane vehicle respectively, while $\Phi(\mathbf{\Xi})$ represents the lane change feasibility function that evaluates the overall safety and practicality of executing a lane change maneuver.

\subsection{Quintic Polynomial Formulation}
For smooth trajectory generation, we employ a quintic polynomial:
\begin{equation}
    y(t) = \sum_{i=0}^5 a_i t^i = a_0 + a_1t + a_2t^2 + a_3t^3 + a_4t^4 + a_5t^5
\end{equation}
where $a_i$ are the polynomial coefficients determining the shape of the trajectory, and $t$ represents time.

The complete trajectory state is described by:
\begin{equation}
    \begin{aligned}
    y(t) &= a_0 + a_1t + a_2t^2 + a_3t^3 + a_4t^4 + a_5t^5 \\
    \dot{y}(t) &= a_1 + 2a_2t + 3a_3t^2 + 4a_4t^3 + 5a_5t^4 \\
    \ddot{y}(t) &= 2a_2 + 6a_3t + 12a_4t^2 + 20a_5t^3
    \end{aligned}
\end{equation}
where $y(t)$ is the lateral position, $\dot{y}(t)$ is the lateral velocity, and $\ddot{y}(t)$ is the lateral acceleration.

The boundary conditions are expressed in matrix form:
\begin{equation}
    \mathbf{M}(t_s, t_e)\mathbf{a} = \mathbf{b}
\end{equation}
where $t_s$ is the initial time, $t_e$ is the final time, $\mathbf{a}$ is the coefficient vector, and $\mathbf{b}$ contains the boundary conditions, with:
\begin{equation}
    \mathbf{M}(t_s, t_e) = 
    \begin{bmatrix}
    1 & t_s & t_s^2 & t_s^3 & t_s^4 & t_s^5 \\
    0 & 1 & 2t_s & 3t_s^2 & 4t_s^3 & 5t_s^4 \\
    0 & 0 & 2 & 6t_s & 12t_s^2 & 20t_s^3 \\
    1 & t_e & t_e^2 & t_e^3 & t_e^4 & t_e^5 \\
    0 & 1 & 2t_e & 3t_e^2 & 4t_e^3 & 5t_e^4 \\
    0 & 0 & 2 & 6t_e & 12t_e^2 & 20t_e^3
    \end{bmatrix}
\end{equation}

\subsection{Vehicle Dynamic Constraints}
The trajectory must satisfy:
\begin{equation}
    \begin{aligned}
    |\ddot{y}(t)| &\leq a_y^\text{max} = 0.4\,g, \\
    |\dot{\psi}(t)| &\leq \frac{\mu\,g}{v_x(t)}, \\
    |\beta(t)| &\leq \beta_\text{max} = 10^\circ, \\
    |\delta(t)| &\leq \delta_\text{max} = 2^\circ.
    \end{aligned}
\end{equation}
where $\mu$ is the road friction coefficient, $g$ is gravitational acceleration, $\beta$ is the side-slip angle, $\delta$ is the steering angle, $\dot{\psi}$ is the yaw rate, and $v_x$ is the longitudinal velocity.

\section{Improved Particle Swarm Optimization}

\subsection{IPSO Algorithm Structure}
The adaptive parameters are defined as:
\begin{equation}
    \begin{aligned}
    w(t) &= w_\text{max} - (w_\text{max}-w_\text{min})\left(\frac{2t}{T}-\frac{t^2}{T^2}\right) \\
    c_1(t) &= c_\text{1,start} + (c_\text{1,end}-c_\text{1,start})\frac{t}{T} \\
    c_2(t) &= c_\text{2,start} + (c_\text{2,end}-c_\text{2,start})\frac{t}{T}
    \end{aligned}
\end{equation}
where $w(t)$ is the adaptive inertia weight controlling exploration vs exploitation,
$c_1(t)$ is the cognitive learning factor influencing personal best attraction,
$c_2(t)$ is the social learning factor influencing global best attraction,
$T$ is the total number of iterations,
$w_\text{max}$ and $w_\text{min}$ are the maximum and minimum inertia weights,
$c_\text{1,start}$, $c_\text{1,end}$, $c_\text{2,start}$, and $c_\text{2,end}$ are the initial and final learning factors.

\subsection{Particle Update Rules}
The particle updates follow:
\begin{equation}
    \begin{aligned}
    v_{i+1} &= w(t)v_i + c_1(t)r_1(p_\text{best}-x_i) + c_2(t)r_2(g_\text{best}-x_i) \\
    x_{i+1} &= x_i + v_{i+1}
    \end{aligned}
\end{equation}
where $v_i$ is the particle velocity, $x_i$ is the particle position, $p_\text{best}$ is the personal best position, $g_\text{best}$ is the global best position, and $r_1,r_2$ are random numbers in [0,1].

The velocity constraints are:
\begin{equation}
    v_\text{min} \leq v_{i+1} \leq v_\text{max}
\end{equation}
where $v_\text{min}$ and $v_\text{max}$ are the minimum and maximum allowed velocities.

\subsection{Cost Function}
The multi-objective cost function is:
\begin{equation}
    J = \sum_{i=1}^5 w_i J_i
\end{equation}
where $w_i$ are the weights for each objective, and:
\begin{equation}
\begin{aligned}
J_1 &= \frac{\max(x_\text{te})}{x_\text{te,max}}, \quad
J_2 = \frac{\max(a_y)}{a_\text{y,max}}, \quad
J_3 = \frac{\max(\dot{\psi})}{\dot{\psi}_\text{max}},\\[10pt]
J_4 &= \frac{\max(\beta)}{\beta_\text{max}}, \quad
J_5 = \frac{\max(\delta)}{\delta_\text{max}}
\end{aligned}
\end{equation}
where $x_\text{te}$ is the terminal position, $a_y$ is the lateral acceleration, $\dot{\psi}$ is the yaw rate, $\beta$ is the side-slip angle, and $\delta$ is the steering angle.

\subsection{Constraint Handling}
The penalty function is:
\begin{equation}
    \phi_j = L_j\sum_{i=1}^N \max(0, g_j(x_i))
\end{equation}
where $L_j$ is the adaptive weight, $g_j(x_i)$ is the j-th constraint function, and $N$ is the number of particles.

The adaptive weights are calculated as:
\begin{equation}
    L_j = \frac{\sum_{i=1}^N \max(0, g_j(x_i))}{\sum_{k=1}^m\sum_{i=1}^N \max(0, g_k(x_i))}
\end{equation}
where $m$ is the total number of constraints.

\subsection{Lane Change Triggering Condition}
A lane change is initiated when:
\begin{equation}
    \begin{cases}
    u_b > u_{b,\text{threshold}} \\
    u_c < u_{c,\text{threshold}} \\
    u_a(\text{current}) > u_a(\text{adjacent})
    \end{cases}
\end{equation}

This triggering system combines three critical conditions: the repulsive field \( u_b \) must exceed its threshold \( u_{b,\text{threshold}} \), indicating significant pressure from the front vehicle; the lane change risk field \( u_c \) must be below its threshold \( u_{c,\text{threshold}} \), ensuring safe conditions in the target lane; and the attraction field \( u_a \) must indicate more favorable conditions in the adjacent lane compared to the current lane.

\subsection{Risk Field Evolution}
The total risk field evolves according to:
\begin{equation}
    \frac{d}{dt}U_\text{total} = \frac{\partial U_\text{total}}{\partial t} + \sum_{i} \frac{\partial U_\text{total}}{\partial \mathbf{X}_i}\dot{\mathbf{X}}_i
\end{equation}

This dynamic equation describes how the total risk field changes over time, where $\frac{\partial U_\text{total}}{\partial t}$ represents the explicit time variation of the field, $\frac{\partial U_\text{total}}{\partial \mathbf{X}_i}$ captures the sensitivity of the field to changes in each vehicle's state, and $\dot{\mathbf{X}}_i$ represents the rate of change of each vehicle's state variables.

\section{Simulation Results}
\begin{figure}[h]
    \centering
    \includegraphics[width=0.6\linewidth]{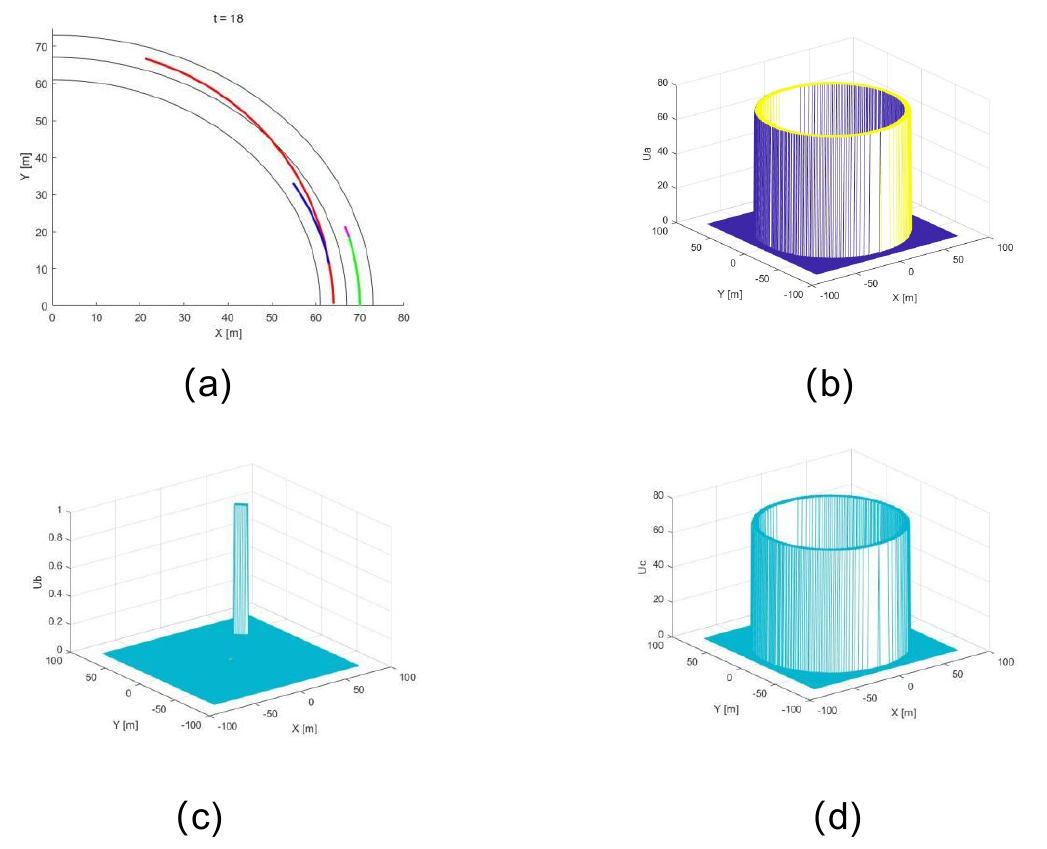}
    \caption{Results of the decision-making for Case 1: (a) the trajectories of the SV and the PV, (b) \( U_a \) at the moment of lane changing, (c) \( U_b \) at the moment of lane changing, (d) \( U_c \) at the moment of lane changing.}
    \label{fig1_framework}
\end{figure}
\begin{figure}[t]
    \centering
    \includegraphics[width=0.6\linewidth]{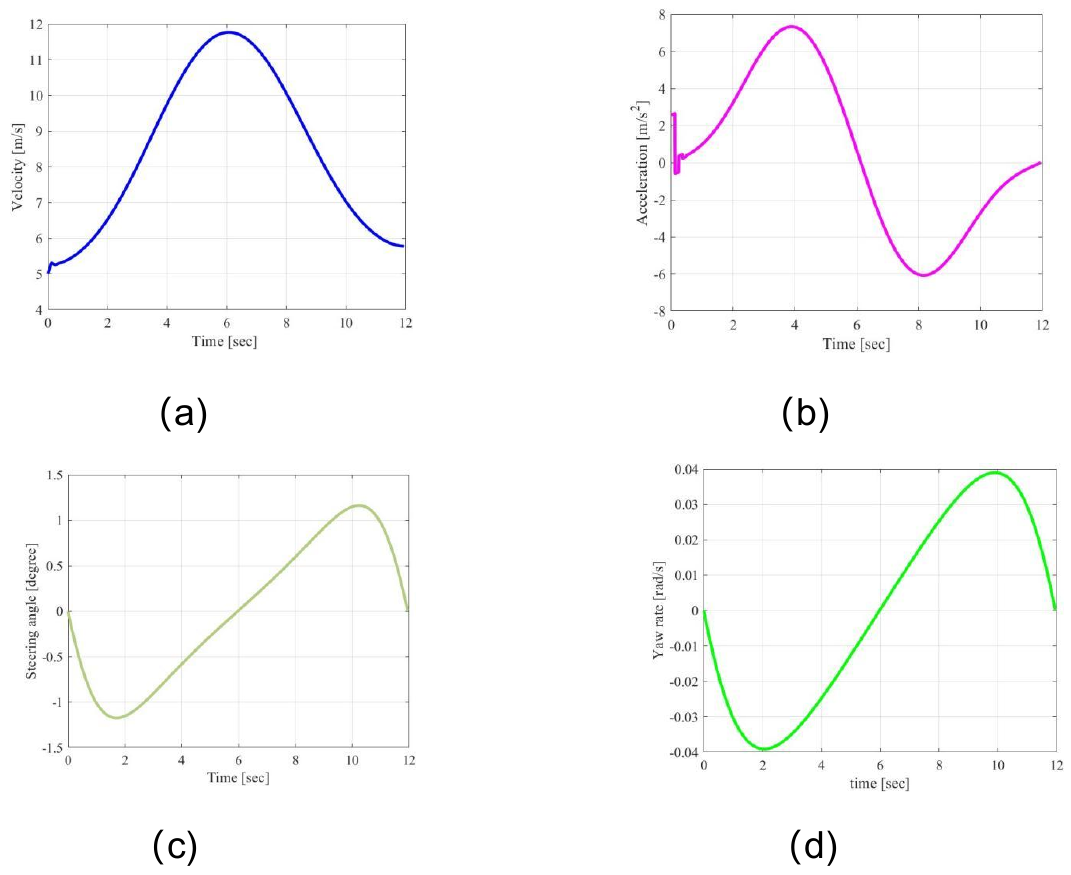}
    \caption{Results of the lane-changing for Case 1: (a) the curve of velocity during lane changing, (b) the curve of acceleration during lane changing, (c) the curve of steering angle during lane changing, and (d) the curve of yaw rate during lane changing.}
    \label{fig1_framework}
    \end{figure}
    \begin{figure}[h]
    \centering
    \includegraphics[width=0.6\linewidth]{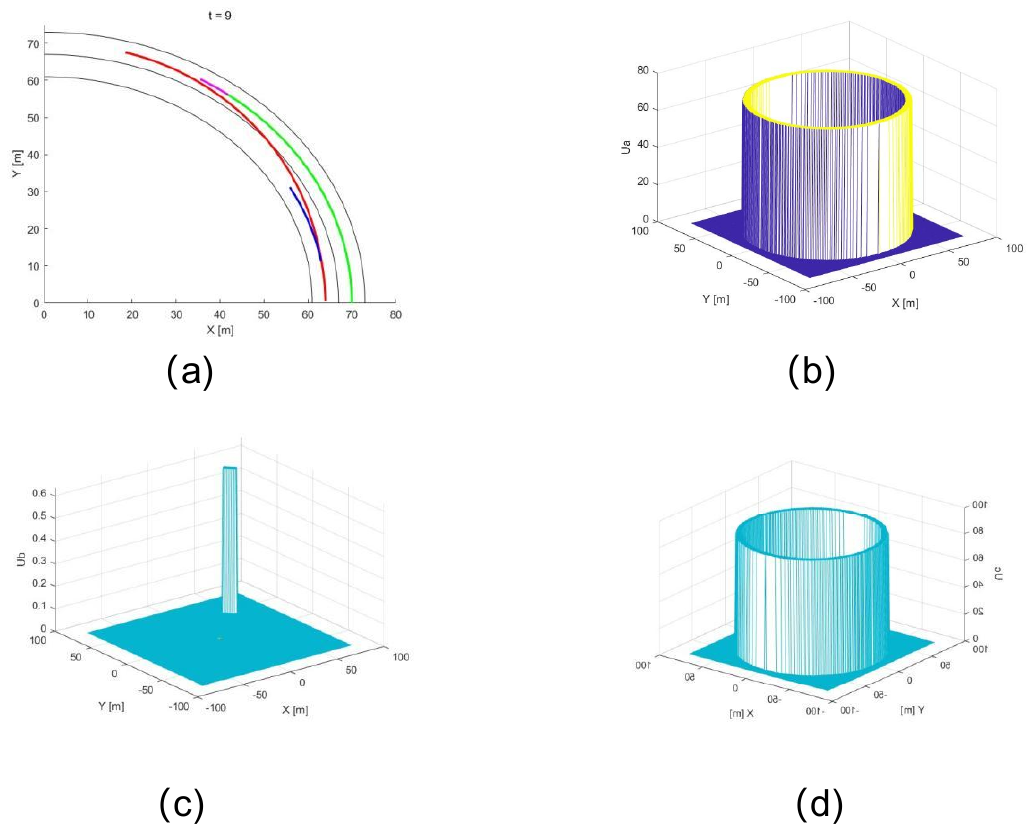}
    \caption{Results of the decision-making for Case 2: (a) the trajectories of the SV and the PV, (b) \( U_a \) at the moment of lane changing, (c) \( U_b \) at the moment of lane changing, (d) \( U_c \) at the moment of lane changing.}
    \label{fig1_framework}
\end{figure}
\begin{figure}[t]
    \centering
    \includegraphics[width=0.6\linewidth]{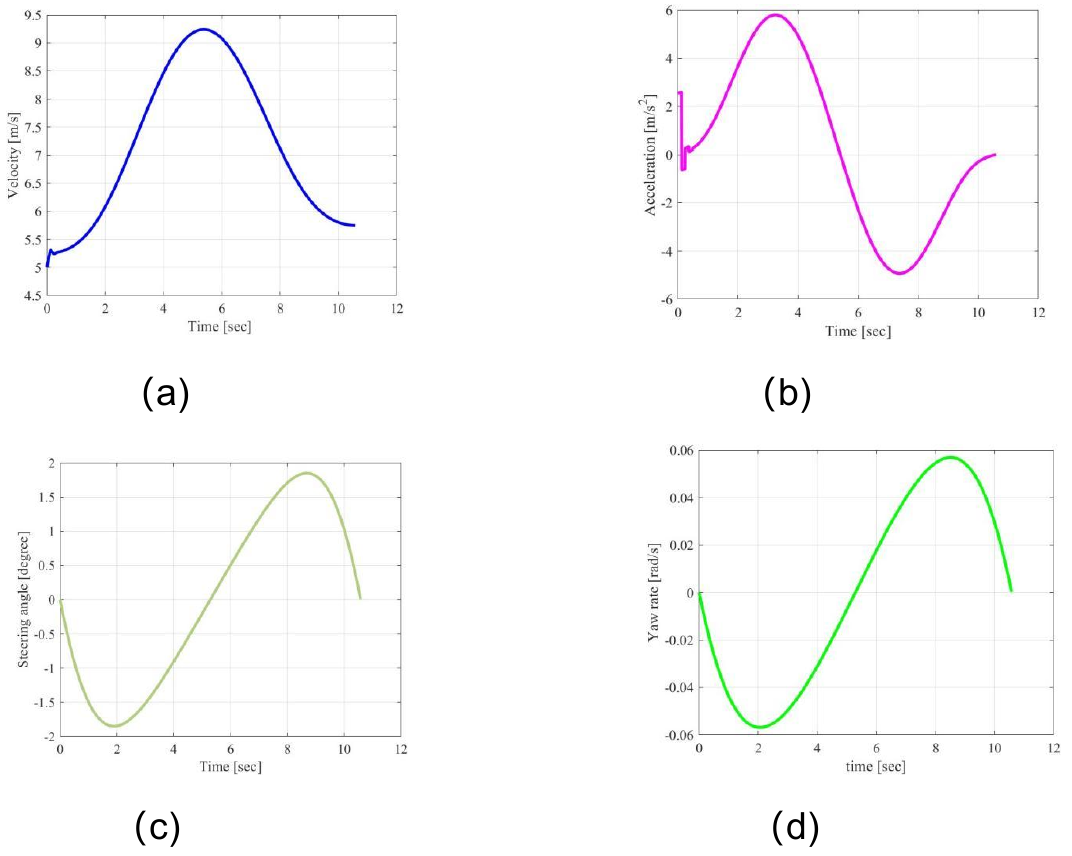}
    \caption{Results of the lane-changing for Case 2: (a) the curve of velocity during lane changing, (b) the curve of acceleration during lane changing, (c) the curve of steering angle during lane changing, and (d) the curve of yaw rate during lane changing.}
    \label{fig1_framework}
    \end{figure}
This section presents our experimental results and an analysis of the proposed framework in Matlab 2022b.
The simulations aim to assess the proposed framework by addressing interactive driving with different iniital settings and HDVs on a curvy road. The curvy road has a inner radium of 64 meters and a outer radium of 70 meters. To verify the generalization of the proposed method, we test SV, PV, RV, and IV with different initial speeds and use the following parameters: initial lane position (inner: 2, outer: 1), initial angle to the parallel line from the center of the circle to Exit 2, and initial speed to represent the initial state of each vehicle. The test cases are:
\begin{enumerate}
    \item SV(2,0,5), PV(2,10,2), IV(1,2,1.5),RV(1,-6,3.5),
    \item SV(2,0,5), PV(2,10,2), IV(1,2,12.5), RV(1,-6,6),
\end{enumerate}

Figs.~1(a) through (d) show the smooth curves of velocity, acceleration, steering angle, and yaw rate, reflecting the comfort during the lane-changing process. Fig.~2(a) illustrates the trajectories of four vehicles. The lane-changing time is 18, which is earlier than in the previous cases. This is due to the higher speed of the RV compared to the IV, which reduces the space of S2 and increases the repulsive field generated by S2. The earlier lane-changing time demonstrates the adaptability of the proposed framework. Comparing Fig.~2(b) to Fig.~2(d), it is evident that the main repulsive field originates from $U_c$, with a value of around 80. Figs.~3(a) through 3(d) present the smooth curves of velocity, acceleration, steering angle, and yaw rate, reflecting the comfort during the lane-changing process. Fig.~4(a) shows that the SV initiates a lane change at time step 9. The early lane change is prompted by the rapid decrease in S2 and the need for the SV to maintain a safe distance from the IV. As a result of the early lane change, the SV successfully reaches the target lane and maintains a safe distance from the IV, ensuring both safe and efficient driving.

\begin{figure}[t]
    \centering
    \includegraphics[width=0.8\linewidth]{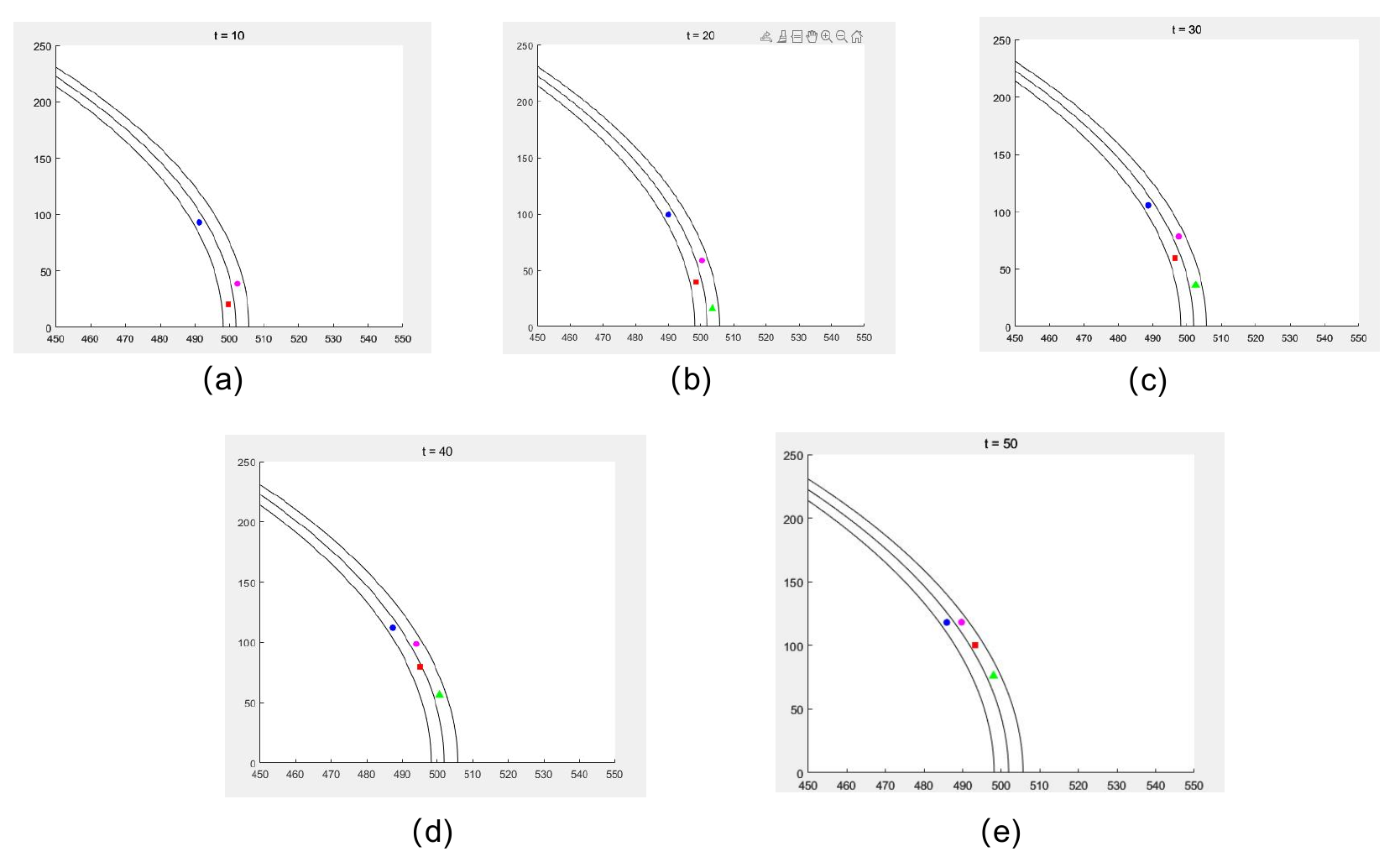}
    \vspace{-2mm}
    \caption{Results of the decision making and lane-changing in a wider curvy road}
    \label{fig1_framework}
    \end{figure}

Fig. 5 illustrates the lane-changing process over discrete time point: the SV in red, the PV in blue, the IV in purple, and the RV in green. Initially, at \( t = 10 \), all vehicles maintain their respective positions in their lanes. By \( t = 20 \), SV moves closer to PV, while IV and RV adjust slightly within their lanes, reflecting dynamic interactions. At \( t = 30 \), SV begins the lane-changing process, preparing to shift to the adjacent lane as PV continues ahead. By \( t = 40 \), SV is actively moving into the adjacent lane, maintaining safe distances from IV and RV. Finally, at \( t = 50 \), SV completes the lane change and establishes its position in the new lane.

 \begin{figure}[h]
    \centering
    \includegraphics[width=0.58\linewidth]{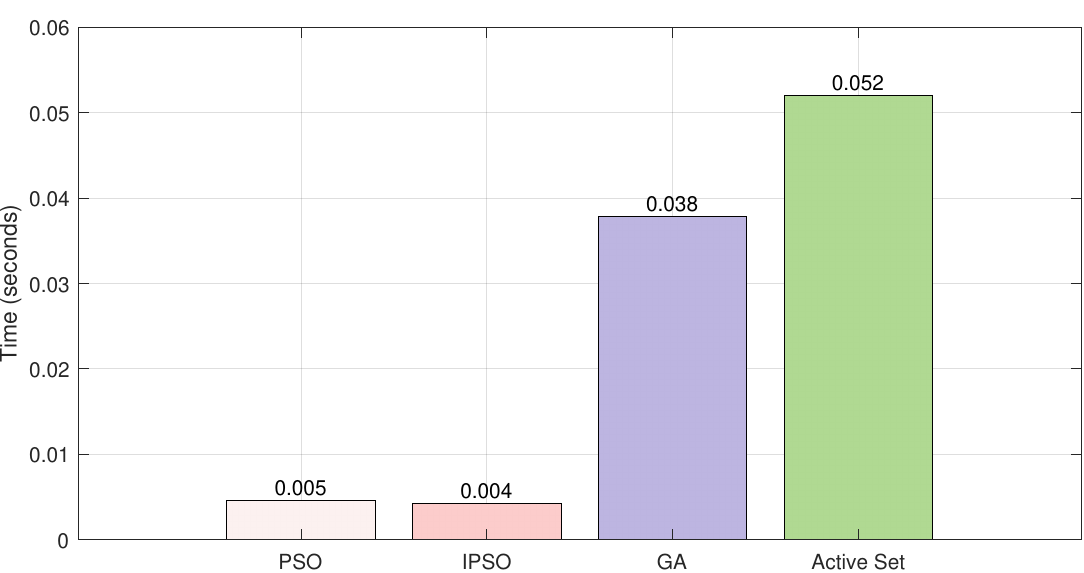}
\caption{Comparison of computational time to convergence for Case 3 across different optimization algorithms}
    \label{fig1_framework}
    \end{figure}
Fig. 6 clearly demonstrates the superior efficiency of the IPSO method compared to other popular optimization methods, including Traditional PSO, Genetic Algorithm (GA) \cite{9312466}, and Active Set methods~\cite{rontsis2022active}. IPSO exhibits the fastest convergence time, requiring only 0.004 seconds on average, which is a significant improvement over Traditional PSO of 0.005 seconds. While both PSO-based methods outperform GA of 0.038 seconds and the Active Set method of 0.052 seconds, IPSO’s reduced computation load highlights its optimized parameter tuning and enhanced convergence mechanisms..

\section{conclusion}
In this paper, we propose a framework that integrates dynamic risk assessment, trajectory generation, and adaptive optimization for interactive driving in complex traffic environments. By employing a novel risk field, vehicles assess safety and efficiency to adjust lane-change decisions based on interactions with surrounding vehicles. The use of Frenet coordinates simplifies trajectory generation on curved roads, while an adapted quintic polynomial method ensures smooth, comfortable transitions. The improved particle swarm optimization (IPSO) algorithm refines trajectory selection via a cost function, achieving faster convergence than popular benchmarks. Extensive simulations demonstrate that our framework effectively handles dynamic obstacles, navigates curvy roads, and meets stringent safety and comfort requirements. Future work will extend the framework to more complex urban scenarios and incorporate machine learning for enhanced prediction of surrounding vehicle behavior.

\bibliographystyle{IEEEtran}
\bibliography{IEEEabrv,zq_lib}

\end{document}